\documentclass[journal=ancac3,
manuscript=article,
layout=traditional
]{achemso}

\usepackage{graphicx}% Include figure files
\usepackage{dcolumn}% Align table columns on decimal point
\usepackage{bm}% bold math
\usepackage{hyperref}% add hypertext capabilities

\usepackage{siunitx}
\usepackage{mathtools}
\usepackage{cleveref}
\usepackage{enumitem}
\usepackage{multirow}
\usepackage{scalerel,amssymb}
\usepackage{verbatim}
\usepackage[dvipsnames]{xcolor}
\usepackage{lineno}
\usepackage[version=4]{mhchem}

\crefname{figure}{Figure}{Figures} % Standard reference
\Crefname{figure}{Figure}{Figures} % Beginning-of-sentence ref
\crefname{table}{Table}{Tables}
\Crefname{table}{Table}{Tables}
\crefname{equation}{Eq.}{Eq.}
\Crefname{equation}{Equation}{Equations}
\title{Can Neural Networks Learn Nanoscale Friction?}

\author{Mahboubeh Shabani}
    \affiliation{Department of Physics, Shahid Beheshti University, Tehran 1983969411, Iran}
    \affiliation{International School for Advanced Studies (SISSA), Via Bonomea 265, 34136 Trieste, Italy}
    \affiliation{International Centre for Theoretical Physics (ICTP), Strada Costiera 11,34151 Trieste,Italy}
    
\author{Andrea Silva}
    \email{ansilva@sissa.it}
    \affiliation{CNR-IOM, Consiglio Nazionale delle Ricerche - Istituto Officina dei Materiali, c/o SISSA, Via Bonomea 265, 34136 Trieste, Italy}
    \affiliation{International School for Advanced Studies (SISSA), Via Bonomea 265, 34136 Trieste, Italy}

\author{Franco Pellegrini}
    \affiliation{International School for Advanced Studies (SISSA), Via Bonomea 265, 34136 Trieste, Italy}

\author{Jin Wang}
    \affiliation{International School for Advanced Studies (SISSA), Via Bonomea 265, 34136 Trieste, Italy}
    \affiliation{International Centre for Theoretical Physics (ICTP), Strada Costiera 11,34151 Trieste,Italy}
    
\author{Renato Buzio}
    \affiliation{CNR-SPIN, C.so F.M. Perrone 24, 16152, Genova, Italy}

\author{Andrea Gerbi}
    \affiliation{CNR-SPIN, C.so F.M. Perrone 24, 16152, Genova, Italy}

\author{Andrea Vanossi}
    \affiliation{CNR-IOM, Consiglio Nazionale delle Ricerche - Istituto Officina dei Materiali, c/o SISSA, Via Bonomea 265, 34136 Trieste, Italy}
    \affiliation{International School for Advanced Studies (SISSA), Via Bonomea 265, 34136 Trieste, Italy}

\author{Ali Sadeghi}
    \affiliation{Department of Physics, Shahid Beheshti University, Tehran 1983969411, Iran}
    
\author{Erio Tosatti}
  \email{tosatti@sissa.it}
  \affiliation{International School for Advanced Studies (SISSA), Via Bonomea 265, 34136 Trieste, Italy}
  \affiliation{International Centre for Theoretical Physics (ICTP), Strada Costiera 11,34151 Trieste,Italy}
  \affiliation{CNR-IOM, Consiglio Nazionale delle Ricerche - Istituto Officina dei Materiali, c/o SISSA, Via Bonomea 265, 34136 Trieste, Italy}

\begin{document}

\begin{abstract}
%AS (9/12/24) word count: 243/250
Current nanofriction experiments on crystals, both tip-on-surface and surface-on-surface, provide  force traces as their sole output, typically exhibiting atomic size stick-slip oscillations. Physically interpreting these traces is a task left to the researcher. Historically done by hand, it generally consists in identifying the parameters of a Prandtl-Tomlinson (PT) model that best reproduces these traces. This  procedure is both work-intensive and quite uncertain. We explore in this work how machine learning (ML) could be harnessed to do that job  with optimal  results, and minimal human work. A set of synthetic force traces is produced by  PT model simulations covering a large span of parameters, and a  simple neural network (NN) perceptron is trained with it. Once this trained NN is fed with experimental force traces, it will ideally output the PT parameters that best approximate them.  By following this  route step by step, we encountered and solved a variety of problems which proved most instructive and revealing. In particular, and very importantly, we met unexpected inaccuracies with which  one or another parameter was learned  by the NN.  The problem, we then show, could be eliminated by proper manipulations and augmentations operated on  the training  force traces, and that without extra efforts and without injecting experimental informations.  Direct application to the sliding of  a graphene coated  AFM tip on a variety of 2D materials substrates validates and encourages use  of this ML  method as a ready tool to rationalise and interpret future stick-slip nanofriction data.
\end{abstract}

\maketitle

% Adhering to ACS guidelines: no Introduction
%\section{Introduction}

\begin{figure*}[!ht]
\centering
\includegraphics[width=0.9\linewidth]{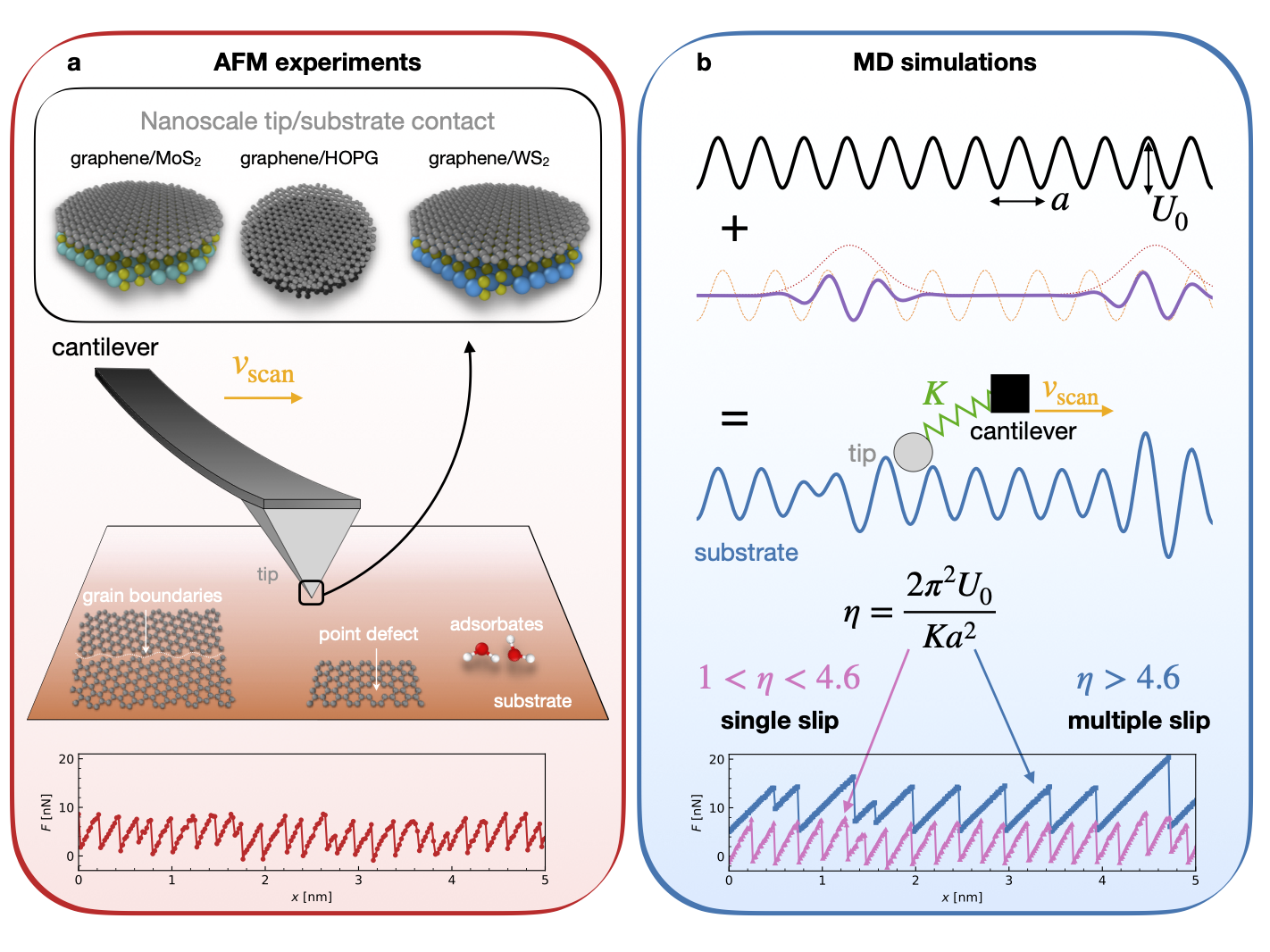}
\caption{
Experimental setup and physical model.
(a) In the main middle panel, sketch of a typical AFM experiment highlighting the moving cantilever and the tip sliding over a substrate at constant speed $v_\mathrm{scan}$.
A set of defects like grain boundaries, point defects and adsorbate, is shown.
The upper inset schematizes the nanoscale contacts realised in our experimental setup, where 2D materials are slide against each other.
The bottom panel reports a representative experimental force trace recorded as the cantilever slides at constant speed $v_\mathrm{scan}$ over the surface.
(b) Sketch of the augmented PT model deployed in this study. 
The model is composed of the standard sinusoidal modulation of height $U_0$ and spacing $a$ augmented with a localised distortion emulating the defects present on the real surface. 
The resulting substrate is depicted in blue, along with the modeled AFM point-like top attached to a translating cantilever by an effective spring of constant $K$. 
The sliding behaviour of the model is determined by the dimensionless parameter $\eta_\mathrm{PT}$.
The simulated force traces relative to the signle and multiple slips regimes are reported at the bottom of panel b in pink triangles and blue squares, respectively. 
}
\label{fig:sys_setup}
\end{figure*}

Friction plays an important role in our daily life. 
An ubiquitous dry friction phenomenon is stick-slip, observed from nanoscale contact shearing to geophysical scale earthquakes. 
This non-equilibrium process, resulting from a mechanical instability first described by Prandtl\cite{prandtl_uber_1905}, is generally highly nonlinear, characterized by long periods of quiet stress accumulation (sticking) followed by sudden and severe energy dissipation events (slips)\cite{vanossi_driven_2007,vanossi_colloquium_2013,wang_colloquium_2024}.
Unfortunately for theorists, the frictional shear between solids, even in its simplest form without lubricants or wear, is  physically complex, involving too many parameters. 
These aspects, non-equilibrium, non-linearity and complexity, make it challenging to describe friction with the tools of statistical physics and to link experimental measures and theoretical predictions.

The usefulness of Machine Learning (ML) is currently emerging across all  disciplines, including some recent applications to tribology\cite{prost_classification_2023,brase_generalised_2024,ying_effect_2024,ying_chemifric_2024,sattari_baboukani_prediction_2020,zaidan_mixture_2017,barik_frictional_2024}.
In this work we show that ML can be harnessed to partly tame and bypass the complexity of stick-slip friction. 
The starting point is the recent simulation\cite{wang_effective_2024} and experimental\cite{buzio_dissipation_2023} work suggesting that for a given velocity and temperature the frictional behaviour of even relatively large meso- and micro-scale sliders can be {\it quantitatively} described by four parameters of an effective Prandtl-Tomlinson (PT) point-slider model. 
In that historical model, so far only validated for nanoscale sliders\cite{socoliuc_transition_2004}, a point  mass $M$ is forced by a spring $K$ to slide over a sinusoidal potential of amplitude $U_0$ (the barrier), the frictional  work absorbed by a damping $\gamma$. 
As a general functional interpolator, ML appears ideally suited to the task of automatically  fitting the experimental frictional data so as to extract the best effective PT parameters, a heavy procedure now carried out with laborious algorithms by experimentalists\cite{buzio_dissipation_2023}.

Here, we therefore address the basic question: can ML capture the physics of nanoscale sliding and yield quantitative prediction for the key PT parameters that control  stick-slip frictional dissipation?
In particular, we  shall focus on Atomic Force Microscopy (AFM) experiments based on the relative sliding of 2D material flakes, as sketched in \cref{fig:sys_setup}. 

Compared to the ideal PT model, the energy landscape of real AFM measurements, sketched in \cref{fig:sys_setup}a,  is complicated by the extended size of the frictional contact, of generally unknown atomic structure and shape, as well as by surface defects including grain boundaries, point defects and adsorbate molecules. 
The friction of extended sliding contacts is also influenced by edges, that constitute omnipresent defects\cite{wang_effective_2024,wang_colloquium_2024,panizon_frictionless_2023}.
In order to capture the essential PT parameters out of this complex energy landscape, we augment the standard PT model sinusoidal potential structure with a long-wavelength modulation mimicking the structural defect and complexity of the real surface\cite{gnecco_friction_2022,gnecco_atomic-scale_2024}. 
The practical goal of this work is to consider raw force traces, like the experimental and simulated ones reported at the bottom of \cref{fig:sys_setup}a,b respectively, and see if a Neural Network (NN) model can learn the effective PT model %system 
parameters from this crude data.

\begin{figure}[!ht]
\centering
\includegraphics[width=0.5\linewidth]{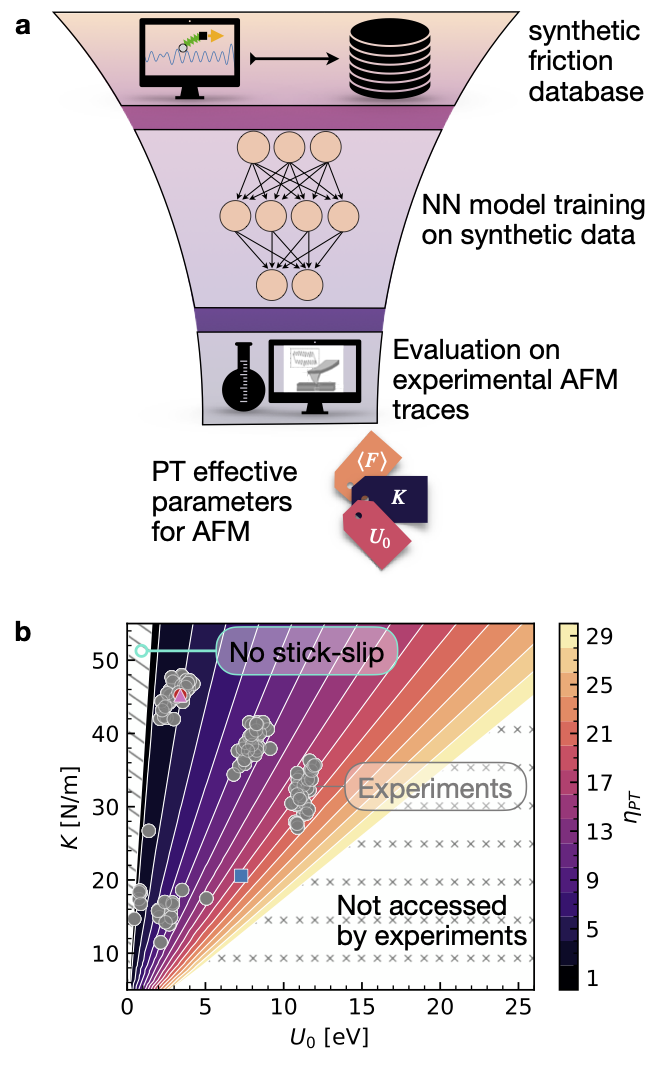}
\caption{
Workflow and system parameter space.
(a) The sketch represent the workflow developed in this study.
Starting at the top, a systematic exploration of the parameter space via MD simulations of the augmented PT model yields an extensive dataset of synthetic force traces.
Proceeding down the workflow, this dataset is used to train a Neural Network (NN) model able to estimate the PT model parameters from force traces.
Finally this synthetic-trained model is used to evaluate real AFM traces to extract effective AFM parameters for the experimental system.
In this study we labeled the AFM traces by an automated algorithm estimating the parameters from the traces and physical evaluations to assess the reliability of the model.
(b) The coloured slices shows the region of relevant parameters $U_0$ and $K$ explored by the simulations.
The color reports the value of dimensionless parameters $\eta_\mathrm{PT}$, see the colorbar on the right.
Gray points mark the values of $U_0$ and $K$ estimated by an automated algorithm in the testing experimental dataset. 
The left (tilted-lines covered) and bottom right (x-covered) patches mark the regions of smooth-sliding and large $\eta_\mathrm{PT}$ not relevant for the considered AFM experiments.
}
\label{fig:param_space_workflow}
\end{figure}

Creating by simulations a curated dataset with the relevant sliding physics is the first step in our workflow, as in \cref{fig:param_space_workflow}a.
This dataset is used to train a NN via supervised learning: a fraction of the raw trajectories are fed to the NN model along with the labels, the remainder is left for validation. 
As we shall discuss, this training yields very reliable results.
The next, more ambitious goal is to then use this model, now trained solely on synthetic data, to identify the best PT model parameters behind all new, raw experimental friction force traces. 
Note that while the training data is generated by a simulated 1D model, the experimental data come from the vastly more complex 2D interfaces between 3D solids. 
The PT model  used is of course unable to capture this complexity. Yet, as we shall see, embedding in the model a few simple additional details makes the problem learnable by the NN,  and the protocol viable.

% ACS guidelines
\section{Results and Discussion}

\begin{figure}[!ht]
\centering
\includegraphics[width=0.5\linewidth]{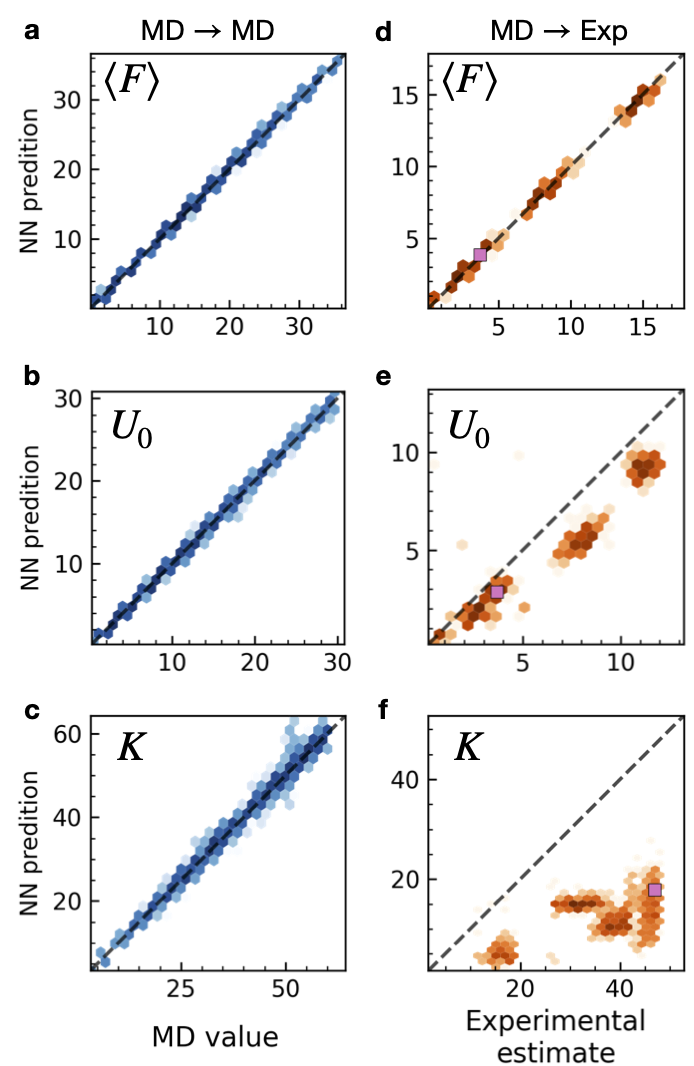}
\caption{
Parity plot for the NN model trained on synthetic data and used to predict other synthetic data (MD$\to$MD left column of panels a-c) or experimental data (MD$\to$Exp right column of panels d-f).
The intensity of the color reflects the local density of points.
Evaluation of the NN in predicting average force $\langle F \rangle$ (a), sliding barrier $U_0$ (b) and effective stiffness $K$ (c) of a test set of simulation that were not shown to the network during training.
Evaluation of the NN in predicting average force (d), sliding barrier (e) and effective stiffness (f) of an experimental dataset of which no element was ever considered during training. 
Note that the prediction power on the effective constant $K$ achieved by the network within the synthetic data-space does not generalise to experimental data.
We consider the average prediction over the full experimental trajectory, not of the single batch.
The pink squares mark the trajectory shown in \cref{fig:physNN_results}g  and analysed in details in \cref{fig:physNN_results}d-f.
}
\label{fig:raw_NN}
\end{figure}

The first step is to define the relevant parameter space for our dataset. 
The main parameters of a PT model suitable for the simulation of an AFM stick-slip force trace are the barrier height $U_0$ and the effective contact stiffness $K$. The mass M and damping $\gamma$ generally take standard values and need not be included in this first step. 
As shown in \cref{fig:param_space_workflow}b, the MD simulations span (coloured regions) all the parameter space relevant for experiments. Our experimental data set  is marked by gray points.
The MD simulations for a broad range of $K$ and $U_0$ values are conducted at room temperature, and with all other parameters, M, $\gamma$ and mean velocity $\langle v\rangle$ fixed as detailed in the Methods section and SI section I.
We focus on the stick-slip regime, including both single slip and multiple slips\cite{medyanik_predictions_2006}.
This procedure yields a simulated database of 1600 trajectories to train the NN model.

We split the dataset in 80/20 between training and validation.
The raw traces accompanied by the $K$ and $U_0$ labels are used in the supervised training of a simple multilayer perceptron with two hidden layers, taking a fixed number of consecutive force measurements as input (see the Methods section for details).
The parity plots for the validation set, showing the real MD parameters against NN predictions, are reported in \cref{fig:raw_NN}a,b,c. 
The plots correspond to the average force, barrier height and effective stiffness, respectively. 
The corresponding RMSEs are reported in \cref{tab:RMSE}.
Clearly, the NN is able to learn the parameters of the underlying model from the simulations to a near-perfect degree.
This is not surprising, as the validation set is part of a coherent dataset of clean data all taken from the same statistical distribution. Moreover, the region of the parameters space from which both training and validation sets come from is well sampled.

\begin{table}[]
    \centering
    \begin{tabular}{l|c|c|r}
        \textit{Model / RMSE} & $\langle F \rangle$ [nN] & $U_0$ [eV] & $K$ [N/m]   \\
        \hline
        NN MD$\to$MD     & 0.16 & 0.18 &  0.84 \\
        NN MD$\to$Exp    & 0.31 & 1.78 & 26.62 \\
        \hline
        NN Exp$\to$Exp   & 0.13 & 0.68 &  2.97 \\
        NN Exp$\to$MD    & 0.42 & 1.54 & 23.33 \\
        \hline
        \textbf{PI-NN MD $\to$Exp} & \textbf{0.18} & \textbf{1.33} & \textbf{4.33} \\
    \end{tabular}
    \caption{RMSE of the same NN model trained on different dataset and different descriptors. 
    NN refers to the model trained on raw force traces, PI-NN to models trained on force traces and processed derivatives. 
    MD and Exp refer to synthetic and experimental datasets.
    The one left of the arrow is the training set, the one right of it is the evaluation set. 
    Note that the two sets are disjoints: a NN trained on the MD dataset is never shown an Exp force trace during training.
    }
    \label{tab:RMSE}
\end{table}

To assess whether the physical contents embedded in the model trained on synthetic data can be recognized in actual frictional stick-slip force traces, we submit to the NN our own dataset of 1298 AFM experimental trajectories, akin to the one shown at the bottom of \cref{fig:sys_setup}a.
The general setup of an AFM experimental system is sketched in \cref{fig:sys_setup}a.
While in the rest of the work we will deal with our own realization of the AFM experiments involving colloidal probes, we note that no assumption link to the specific setup are made in our treatment and, thus, we believe our results are relevant for any AFM experiment observing atomic stick-slip\cite{liu_interlayer_2018,buzio_subnanometer_2018,dasic_role_2024,vazirisereshk_nanoscale_2020,yang_atomic_2024,song_velocity_2022,marian_physics-informed_2023,vilhena_atomic-scale_2016}. 
Our AFM colloidal tip is capped with a nanorough graphene coating  of mesoscopic size, which ultimately controls the contact mechanics and friction through tens-of-nanometers tall protrusions which slide at ambient conditions on the substrate. 
Substrates were either graphite (HOPG) or TMDs with a crystalline axis parallel to the tip  sliding direction\cite{buzio_dissipation_2023,buzio_graphite_2021,buzio_sliding_2022}.
Unlike the perfect sketch of \cref{fig:sys_setup}a the contact itself will inevitably contain structural defects and adsorbates. (See Methods and SI section II for details.)
The putative PT parameters that best describe the experimental force traces in our experimental setup were in fact estimated manually in previous work, with an \textit{ad hoc} algorithm\cite{socoliuc_transition_2004,jinesh_thermolubricity_2008,buzio_graphite_2021,schirmeisen_tip-jump_2005}.
That and other works indeed suggested that the PT model, generally used for point-like sliders, could capture well the frictional physics of sliding contacts up to mesoscopic size\cite{vanossi_driven_2007,wang_effective_2024}.

As reported in \cref{fig:raw_NN}d,e,f the learning of average friction $\langle F \rangle$ and barrier $U_0$ extrapolates well from synthetic trajectories to experimental ones, while the extrapolation of $K$ fails: the predicted values are scattered almost randomly.
What is the origin of this failure?
First, we did a necessary sanity check. 
The culprit is not simply the fact that the experimental traces do not carry enough information for the NN model to learn. 
Indeed, training the model on experimental trajectories can relatively well predict the parameters of other experimental trajectories, see SI sec III and \cref{tab:RMSE}.
Moreover, the fault lies not with the \textit{ad hoc} algorithm either. 
If this algorithm is used to estimate the PT parameter from synthetic traces, one n fact recovers the known input parameter with  quite good accuracy (see SI Fig. S3 c-d). The small deviations actually resemble those one would find if the parameters $K$ and $U_0$ were hypothetically extracted by hand from the simulated force traces-- the latter an  operation which makes no further sense in our context.

In fact, the step we are attempting is less trivial than one may suppose from the similarities between force traces in \cref{fig:sys_setup}.  
The synthetic and experimental force traces originate from different systems and are thus drawn from a different statistical distribution.
As the training is done only on the synthetic trajectories, the NN is not penalized for predicting poorly the effective parameters that best describe the experimental force traces.
When the network is trained only on synthetic data, it may in fact become overly specialized in recognizing features unique to this dataset rather then more general, physics-based fingerprints in the traces.
Learning such fine features would permit a more accurate extraction of the best effective PT parameters from the experimental force traces.

In order to test this hypothesis, we mixed a fraction of experimental trajectories in the training set. 
As shown in SI section III, the addition of a small fraction of 1\% of experimental  data in the training set helps the NN to generalize better.
As it exposes the model to the experimental data distribution, with its real-world variability and noise, the mixed dataset encourages the model to focus on general, physics-based features rather than features specific of the synthetic data.
Note that the reverse reasoning works as well: a model trained solely on experimental data fails to predict the synthetic data correctly, see SI III and \cref{tab:RMSE}.

That said, we still need to make progress on a {\it  scheme where the training is restricted exclusively on synthetic data.}
As we shall see in the following section it is possible to enhance the perception of physically meaningful parameters by elaborating the synthetic training data in a way that nudges the NN's learning process in the right direction.

\subsection{Physics-Informed Data Augmentation}

\begin{figure}[!ht]
\centering
\includegraphics[width=0.5\linewidth]{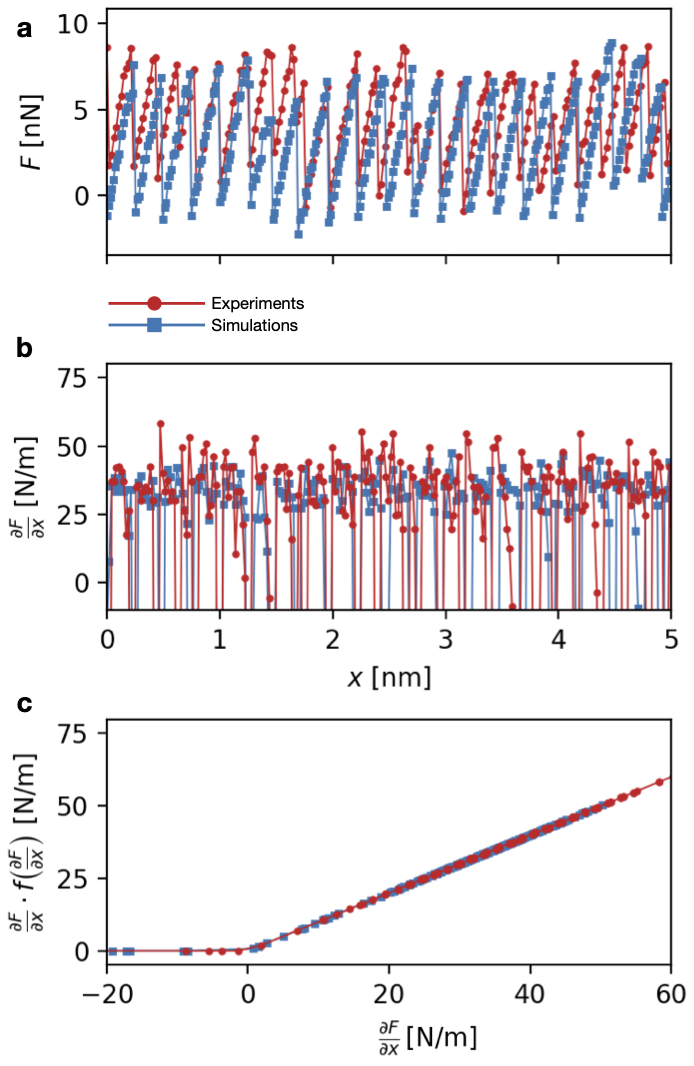}
\caption{
(a) Example of force traces feed to the NN during training. Red and blue symbols refer to an experimental and simulated trace (respectively) with similar parameters. 
(b) Derivative of the force traces in panel a. 
(c) Processed derivative feed to NN during training.
}
\label{fig:physNN_traces}
\end{figure}

We can use our own knowledge of how the physics of the PT model reflects on the force traces to guide the NN training.
We start by noticing that the NN can predict well the barrier height $U_0$, see \cref{fig:raw_NN}e. 
Consider the synthetic and experimental force traces reported in \cref{fig:physNN_traces}.

The barrier is a ``global" quantity in the force trace,
proportional to the maximum of the recorded force\cite{socoliuc_transition_2004,jinesh_thermolubricity_2008,buzio_graphite_2021} (see SI section II).
On the other hand, the effective stiffness $K$ is a more ``local" quantity: to estimate it from the force trace one has to consider the slope of the trace in the ``sticking" part, when the spring charges before the mechanical slip instability %(slip) 
happens. 

The stiffness is  thus a function of the force derivative,  $K=K(\partial F / \partial x)$.
In order to nudge the learning of the NN model in this direction, we therefore feed to the model, in addition to the force trace, also its derivative  $\partial F / \partial x$, computed through finite differences, see \cref{fig:physNN_traces}b.  
We note in addition that only the sticking portion of the trace, where the force slope is positive, is relevant to stiffness while negative slopes do not and could be set to zero in the NN training.
Considering that in experiments the slopes at the sticking parts tends to  be smaller than the real maximum slope (Fig.~4a) due to multiple factors (e.g., defects, temperature, sliding directions),
one may strategically reduce the influence of these smaller slopes on the NN by decreasing their weight.
To achieve this, the derivative is weighted with a sigmoid function $f(x) = 1/(1+\exp(-(x-\mu)/\sigma))$, where $\mu$ is the mean and $\sigma=15$ is a smearing parameter.
As our simple network has no positional encoding, we simply concatenate the function and its processed derivative as an input.
Hence, we feed the NN the sorted value of this function $\partial F/\partial x \cdot f(\partial F/\partial x)$, yielding the curves reported in \cref{fig:physNN_traces}c.
The proposed augmentations is grounded in the physical principles of the PT model. 
This physics-driven preprocessing step improves the ability of the NN to learn the physical principles that underlie the role $K$ in shaping the force trace and ensures a better generalization to experimental data.

\begin{figure*}[!ht]
\centering
\includegraphics[width=0.8\linewidth]{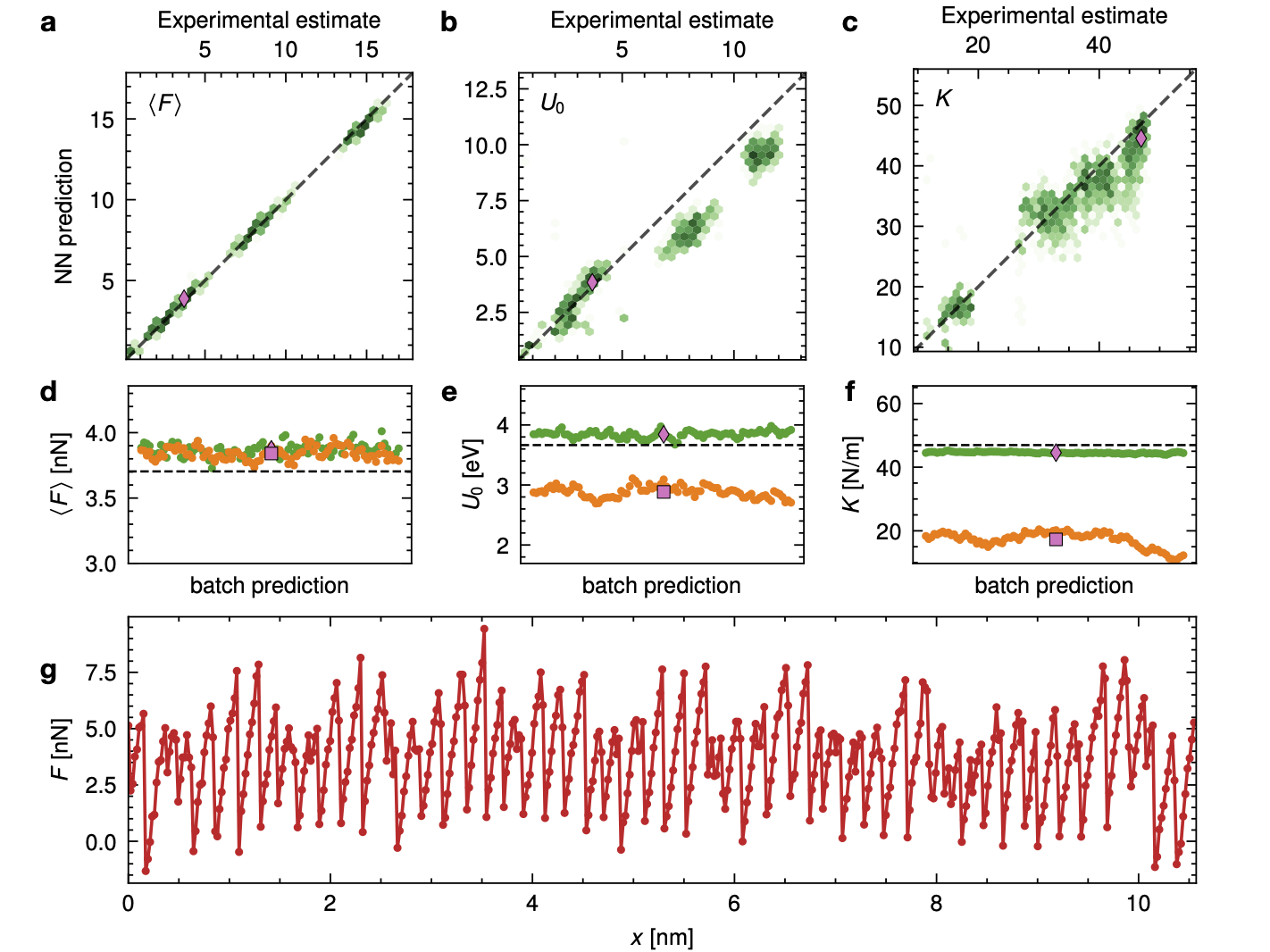}
\caption{
NN trained on physics-augmented synthetic data and evaluated on experimental data.
The color intensity reflects the local density of points.
The NN prediction for (a) average force $\left< F \right>$, (b) sliding barrier $U_0$ and (c) effective stiffness $K$ of an experimental dataset of which no element was ever considered during training. 
Note that the network has now learned to predict $K$ for the experimental data, albeit with a wide spread.
The pink diamonds mark the trajectory shown in panel g and analyzed in details in the panels d-f.
Batch estimates for (d) $\left< F \right>$, (e) $U_0$, and (f) $K$ from the bare NN (orange points) and physics-embedded NN (green points) for the experimental force trace reported in panel (g) (red points). 
This experimental force trace corresponds to the pink diamond in panel a-c and to the pink square in \cref{fig:raw_NN}d-f.
The dashed lines indicate reference values obtained from experimental data using the automated algorithm outlined in the SI.
The averaged prediction for the physics-embedded NN (pink diamond) agrees with the estimated value while the bare NN's prediction is largely underestimated and the batch-wise estimate shows a non-trivial scattering, %hinting 
suggesting that the bare NN has not properly learned how to accurately evaluate $K$.
}
\label{fig:physNN_results}
\end{figure*}

The results of this new procedure are reported in \cref{fig:physNN_results}a,b,c.
Note that now the NN, trained solely on synthetic data, is able to make a reasonable prediction for all the relevant experimental parameters, including the spring constant $K$, see also \cref{tab:RMSE}.
This is remarkable, as the network has never seen any experimental trace. We interpret this as evidence that the NN has learned the key physical mechanism rather than a database-specific pattern and, thus, is able to extrapolate from synthetic traces to experimental ones -- of course within the limit of our crude approximations and model,  which justifies the spread in the predicted points.
A separate origin for the spread is the defects and variability in the experimental samples; indeed training a NN on experimental data to predict other experimental data, results in a similar spread (see SI section III and \cref{tab:RMSE}), hinting at the higher noise intrinsic to the experimental dataset.

The augmented synthetic training scheme now including the rectified force derivative outperforms the one trained on raw traces in all predictions, as seen in \cref{tab:RMSE}. 
The batch-wise predictions for the specific experimental trajectory in \cref{fig:physNN_results}g (highlighted by the pink symbols in \cref{fig:raw_NN}d-f and \cref{fig:physNN_traces}a-c) are shown in \cref{fig:physNN_results}d,e,f . 
Comparing cases were predictions for both are poor, these occur when the experimental traces are noisy and irregular, suggesting that the AFM tip is likely crossing structural defects and different domains. 
In such cases even a human estimate of the PT parameters, our reference label here, is not without ambiguity. 
Moreover, as $U_0$ decreases and $K$ increases, the system approaches the smooth sliding regime. The stick-slip fingerprint weakens and, thus, estimating the system parameters from it becomes less and less reliable, becoming totally impossible in a regime of smooth sliding.\\

As a last observation, part of the uncertainty of the parameters $K$ and $U_0$ can be instructively explained on physical grounds.
The PT model stiffness $K$ is that of an ideal elastic spring.  
The effective experimental contact stiffness is considerably more complicated, generally involving dissipative and plastic deformations taking place under practical loading conditions.
As stress builds up towards the end of a sticking interval, the real contact undergoes precursor relaxations before yielding,  thus softening the  friction force slope just prior to slip. 
The NN reads that last-minute softening as a decrease of overall stiffness and barrier below  those far from the slip. 

Besides that precursor softening, the small systematic underestimation of the barrier $U_0$ also depends on our assumption of a graphene lattice spacing in all simulations, whereas the actual experimental input was more varied, including materials with lattice spacing slightly larger that that of graphene. 
The systematic underestimate of $U_0$ is a perfectly reasonable result rooted in that approximation -- necessary in this conceptual work -- rather than NN capabilities (see Si Fig. S5b and Fig. S6e). 
In fact, this result highlights the considerable robustness of our ML approach.

\section{Conclusions}
We have shown that NN can learn nanoscale friction within the framework of the PT model. 
In order to extrapolate from synthetic, simulated data to real experimental data, care must be exercised: a model too specialized to a single-origin dataset may fail spectacularly. 
To prevent this, the NN training should be constructed so as to include enough physically relevant descriptors.
In this exploratory work we limited ourselves to a relative simple NN architecture according to the fundamental character of our investigation. 
Our aim was to show that indeed the PT framework can be learned by a NN model. As a result, we confirm that training on strictly synthetic data can  indeed yield reasonable  predictions for the best parameters that physically describe  completely unknown experimental data. 
The ML scheme can be naturally extended if desirable to involve sliding models richer  than simple PT, with new parameters related to  other local features of the force traces that could be learned, possibly by additional suitable manipulations of the synthetic training data, as was done here for stiffness.
We also believe that deploying a more sophisticated ML model exploiting the sequential nature of the data, such as a recurrent neural network or even an attention-based model, could improve the learning accuracy. 
The scheme outlined in this work should prove of considerable practical use in direct connection with experimental AFM data post-processing, as well as  of physical value in their interpretation.

\section{Methods}
\paragraph{Synthetic friction simulations}
A 4th order Runge Kutta algorithm was used to propagate the (underdamped) Langevin equation in eq 4 in SI section I.
The instantaneous lateral force trace was evaluated as: $F=-K(x - V_0 t)$ to obtain friction force vs cantilever displacement traces.
We used a tip mass of $m_0 = \SI{1e-12}{Kg}$. 
The Langevin damping coefficient was set to $2 \gamma m = 0.01 ~ \mathrm{ns}^{-1}$ to match experimental force traces amplitude. 
To simulate experimental conditions, we selected a sliding velocity of $v = 60 ~\text{nm/s}$ and a temperature of $T = 296 ~ \text{K}$.

\paragraph{Experimental setup}
The atomic-scale friction force spectroscopies were carried out under standard laboratory conditions, by means of a commercial AFM operated in contact mode (Solver P47-PRO by NT-MDT, Russia). 
An additive-free aqueous dispersion of graphene was used to prepare graphene-coated colloidal probes (based on silica beads of $\sim 25\mu m$ diameter), following a fabrication method ''mixing'' dip-coating and drop-casting techniques\cite{buzio_dissipation_2023}.
High-resolution micrographs revealed that the graphene coating was generally inhomogeneous at the sub-micrometric scale, with uncoated silica regions interspersed with tens-of-nanometers tall protrusions formed by randomly stacked and/or highly crumpled flakes. 
Despite their random accumulation and non-conformal adhesion to the silica surface, the agglomerated flakes maintained a lubricious behavior\cite{buzio_dissipation_2023}. 
Consequently, the manifestation of graphene-mediated friction effects was ultimately controlled by the topographically highest contact nanoasperity\cite{buzio_graphite_2021,buzio_sliding_2022}. 
The graphene-coated colloidal probes were placed in contact with the freshly-cleaved surfaces of HOPG (grade ZYB by MikroMasch), $2H$-WS$_{2}$ or $2H$-MoS$_{2}$ crystals (from HQ Graphene) respectively, thus leading to the realization of different sliding interfaces between 2D materials (as shown in Fig.1a in the main text).
For the calibration of the elastic constant of each probe $k_C$, and of the normal force $F_N$ and lateral force $F_L$, see the Supplementary Information section S1 in Ref.~\cite{buzio_dissipation_2023,schirmeisen_tip-jump_2005}. 
We obtained load-dependent atomic-scale stick-slip trajectories from friction maps ($512 \times 512$ pixels), in which $F_{N}$ was systematically decreased every ten lines from a relatively large starting value (i.e. a few hundreds of nN) to the pull-off point. 
We interrogated surface portions that were free from atomic steps, with a typical scan range $11 \times 11 \mathrm{nm}^2$ and sliding velocity $\sim 30 \mathrm{nm/s}$.  
Representative stick-slip trajectories for each system are shown in Fig. S2 in the SI.
The dataset of 1298 AFM experimental trajectories comprised: 459 traces for the graphene/HOPG interface; 429 traces for the graphene/MoS$_2$ interface; 410 traces for the graphene/WS$_2$ interface.

\paragraph{Neural Network Model}
This investigation leverages PyTorch\cite{paszke_pytorch_2019} to manage the dataset and train a multilayer perceptron $\text{(MLP)}$ architecture designed to predict multiple target features related to friction data, including average force $\langle F \rangle$, sliding barrier $U_0$ and effective stiffness $K$. 
The input to the network is a fixed number $p = \text{sample\_size}$ of consecutive samples from the original force trace. For the physics informed model, we concatenate the trace and finite elements processed derivative for each sample, resulting in an input dimension $p = \text{sample\_size} \times 2 - 1 $. 
The MLP consists of two hidden layers with 256 and 128 neurons, respectively, utilizing the Rectified Linear Unit (ReLU) activation function to introduce non-linearity and enhance the model's ability to capture complex patterns. 
The output layer is linear and has size equal to the number of target quantities.
The loss function is the sum of the quadratic loss for each target. 
The model is trained over 50 epochs with a batch size of 300, using the Adam optimizer with a learning rate of $5 \times 10^{-4}$. 

\section{Acknowledgments} 
Work carried out under ERC ULTRADISS Contract No. 834402.
This study was financially supported by the MIUR PRIN2017 project 20178PZCB5 ``UTFROM -- Understanding and tuning friction through nanostructure manipulation''.
We are grateful to E. Meyer and A. Khosravi for discussions. 
M.S. acknowledges hospitality and help by the Abdus Salam  International Centre for Theoretical Physics, and the International School for Advanced Studies during part of this work,  as well as the financial support of Ministry of Science, Research, and Technology of
Iran. A.S. is grateful to A. Trost for the help with the density plots.

\bibliography{biblio}

\end{document}